\documentclass[aps,prl,reprint,amsmath,amssymb]{revtex4-1}

\usepackage{graphicx}
\usepackage{bm}

\newcommand{\Deltat}{\tilde{\Delta}}

\begin{document}

\title{Quantum dynamics of a microwave resonator strongly coupled to a tunnel junction}

\author{J\'er\^ome Est\`eve}
\author{Marco Aprili}
\author{Julien Gabelli}

\affiliation{Laboratoire de Physique des Solides, CNRS, Universit\'e Paris-Sud, Universit\'e Paris-Saclay, Orsay, France}
\date{\today}

\begin{abstract}
We consider the coupling of a single mode microwave resonator to a tunnel junction whose contacts are at thermal equilibrium. We derive the quantum master equation describing the evolution of the resonator field in the strong coupling regime, where the characteristic impedance of the resonator is larger than the quantum of resistance. We first study the case of a normal-insulator-normal junction and show that a dc driven single photon source can be obtained. We then consider the case of a superconductor-insulator-normal and superconductor-insulator-superconductor junction. There, we show that the Lamb shift induced by the junction gives rise to a nonlinear spectrum of the resonator even when the junction induced losses are negligible. We discuss the resulting dynamics and consider possible applications including quantum Zeno dynamics and the realization of a qubit.
\end{abstract}

\maketitle
The coupling of a quantum system, such as an atom or a resonator, to a bath can be modeled under certain hypotheses by a quantum master equation in the Lindblad form \cite{gardiner2004quantum}. In this description, two effects arises from the coupling to the bath: the first is a shift of the energy levels, called Lamb shift, and the second is an irreversible energy exchange with the bath in the form of quantum jumps. In quantum optics, the jumps correspond to the absorption or emission of photons. The resulting nonunitary dynamics has been proposed as an alternative to unitary evolution for certain quantum engineering tasks as first proposed in \cite{Poyatos:1996va}. The implementation of these ideas requires baths that can be experimentally tailored. In the context of circuit quantum electrodynamics, tunnel junctions offer a simple way to realize a tunable bath when coupled to the mode of a microwave resonator \cite{Bergenfeldt:2012vo,Mendes:2015ue,Grimsmo:2016gr}. 

Here, we concentrate on the strong coupling regime, where the characteristic impedance $Z_c$ of the resonator is on the order of the quantum of resistance $R_K=h/e^2$ as recently realized in \cite{Altimiras:2013et,Samkharadze:2016tw,Stockklauser:2017tr,Kuzmin:2018wy,Martinez:2018wp}. Considering the two contacts of the junction as baths at thermal equilibrium, we derive the quantum master equation describing the evolution of the resonator density matrix following the standard quantum optics approach, which uses the Born-Markov and the secular approximations \cite{gardiner2004quantum}. An important consequence of the strong coupling regime is that the matrix elements of both the Lamb shift and the jump operators depend on the resonator state. Different Fock states experience different Lamb shifts, which introduces a nonlinearity, and quantum jumps from a specific Fock state to another can be forbidden by tuning the coupling parameter $\lambda = \sqrt{\pi Z_c/R_K}$ to specific values \cite{Souquet:2016km}. 

We analyze the effect of these state dependent quantum jumps and energy shifts for the three possible types of junctions that consist of two electrodes made of normal (N) or superconducting (S) metal separated by an insulating barrier (I). In the case of a NIN junction, we show that, for $\lambda^2 = 2$, the absorption of a second photon by the resonator is blocked when it is already occupied by one photon, thus realizing a single photon source. In the case of the SIS and SIN junctions, we show that the Lamb shift gives rise to a sizable nonlinearity, as a consequence of the abrupt increase in conductance of the junction for energies larger than the superconducting gap edge. The nonlinearity can be strong without adding any dissipation to the resonator. This Lamb shift engineering appears as an interesting new facet of quantum bath engineering. We show that it can be used to restrict the dynamics of the resonator mode to the subspace spanned by the lowest $n_c$ Fock states, where $n_c$ can be tuned with the dc voltage biasing the junction. We study first the specific case of $n_c=5$ and analyze the resulting quantum Zeno dynamics during which a four photon cat state is produced \cite{Facchi:2009up, Raimond:2012ta}. We then consider $n_c=2$ and realize a qubit that does not rely on the Josephson effect.     

\section{Master equation}   
The considered circuit is shown in figure 1 and consists of a single mode resonator in series with a tunnel junction and a dc voltage source. This circuit has been extensively studied in the case where the junction is a Josephson junction \cite{Padurariu:2012uh,Leppakangas:2013ty,Armour:2013wg,Gramich:2013gb,Dambach:2015wv,Trif:2015wo,Souquet:2016km}, a NIN junction in the weak coupling regime ($\lambda \ll 1$) \cite{Mendes:2015ue,Grimsmo:2016gr,Mora:2017vb} or in the strong coupling regime ($\lambda \simeq 1$), but considering only the dynamics of the junction \cite{Souquet:2014gm} and a SIN junction \cite{Silveri:2017wl}. The case where the junction is replaced by a quantum dot has also been extensively considered (see \cite{Viennot:2016bu} and references therein). Here, we combine these different approaches and treat the junction contacts as baths, which are coupled to the resonator via the tunnel Hamiltonian. Following the standard quantum optics procedure based on the Born-Markov and secular approximations, we derive in the supplementary materials (SM) the master equation in the Lindbald form for the density matrix of the resonator:
\begin{equation}
\frac{d\rho}{dt} = \sum_{l=-\infty}^\infty -i \, \epsilon_l \left[ A_l A_l^\dagger, \rho \right] +  \gamma_l \left( 2 A_l^\dagger \rho A_l -  \{A_l  A_l^\dagger, \rho \}  \right) \label{eq.mastereq} .
\end{equation}
The $A_l$ operators come from the expansion of the displacement operator $\exp[ i \lambda(a + a^\dagger)]$ in the Fock state basis, where $a$ is the annihilation operator for the resonator mode \cite{Souquet:2016km}. The $A_l$ operator creates $l$ photons in the resonator mode when $l>0$ and destroys $-l$ photons when $l<0$. The first term on the rhs of the master equation is the Lamb shift, which amounts to adding $\sum_l \hbar \epsilon_l  A_l A_l^\dagger$ to the resonator Hamiltonian. The second term describes quantum jumps where $l$ photons are removed ($l>0$) or added ($l<0$) to the resonator mode. The rates $\gamma_l$ and the energies $\epsilon_l$ only depend on the properties of the bath (here the type of junction) and are functions of the dc bias voltage $V$. 

The $A_l$ operators are defined for $l\geq 0$ by  
\begin{equation*}
A_{l \geq 0} = \sum_{n \geq 0} | n+l \rangle \langle n+l | e^{i \lambda(a + a^\dagger)} | n \rangle \langle n |.
\end{equation*}
Operators with $l<0$ are obtained using $A_{-l}=(-1)^lA_l^\dagger$. In the weak coupling limit, the master equation can be simplified to keep only terms involving $A_1=i\lambda a^\dagger$ and $A_{-1}=i\lambda a$. The main goal of this paper is to look at the role of higher order $A_l$ operators, which cannot be neglected when $\lambda \sim 1$. For an arbitrary coupling, the matrix element $w_{n,l}(\lambda)=\langle n+l | e^{i \lambda(a + a^\dagger)} | n \rangle$ can be expressed in terms of the generalized Laguerre polynomials $L_n^{(l)}$ as \cite{Cahill:1969cz}
\begin{equation*}
w_{n,l}(\lambda) = \sqrt{\frac{n!}{(n+l)!}} e^{-\lambda^2/2} \, (i \lambda)^l \, L_n^{(l)}(\lambda^2).
\end{equation*}
Coefficients with $n=0$ correspond to processes where $l$ photons are absorbed or created with the vacuum as the final or initial state, they follow the Poisson law $|w_{0,l}|^2 = \exp(-\lambda^2) \lambda^{2l} / l!$ \cite{Beenakker:2001ur}. Figure 1 shows the evolution of a few $w_{n,l}(\lambda)$ as a function of $\lambda$. One notices that the numbers of zeros of $w_{n,l}(\lambda)$ is $n$.   

\begin{figure}
\includegraphics{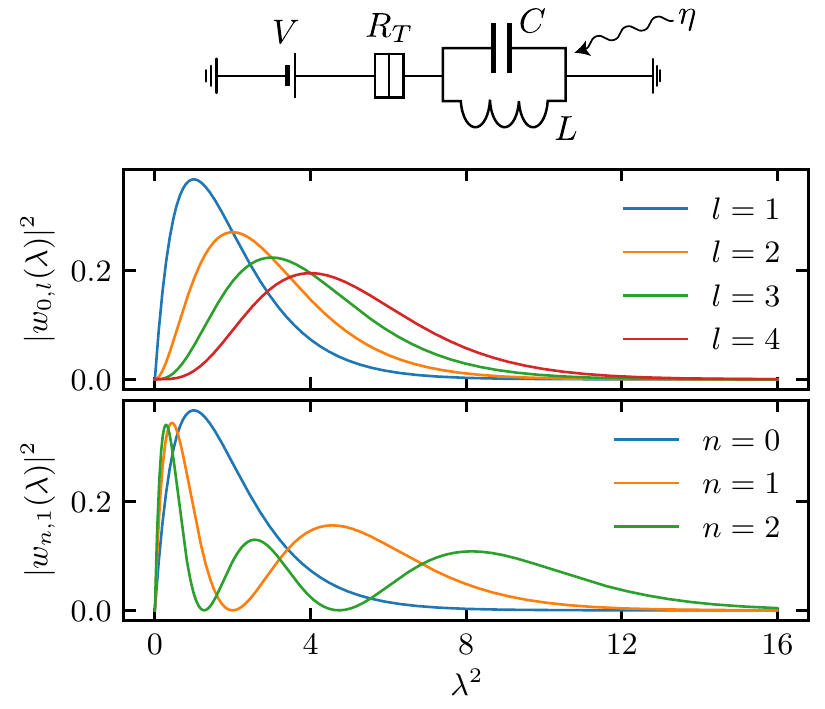}
\caption{We consider a circuit consisting of a $LC$ resonator in series with a tunnel junction and a dc voltage source. The resonator is characterized by its resonant frequency $\Omega=1/\sqrt{LC}$ and its characteristic impedance $Z_c=\sqrt{L/C}$. A coherent drive with amplitude $\eta$ can eventually be applied to the resonator. The effect of the coupling to the junction on the resonator dynamics is conveniently described by introducing $A_l$ operators that add $l$ photons to the resonator mode. The curves in the lower graph show the evolution of the matrix elements $w_{n,l}$ of these operators. In the weak coupling regime ($\lambda \ll 1$), only the effect of $A_1$ and $A_{-1}$ must be considered. In the strong coupling regime ($\lambda \sim 1$), operators with larger $l$ also contribute.}
\end{figure}

The rates $\gamma_l$ and the frequency shifts $\epsilon_l$ are proportional to the real and imaginary parts of the one sided Fourier transform of the bath memory function. As shown in the SM, we obtain
\begin{eqnarray*}
\gamma_l  & = & \gamma(l\Omega +eV/\hbar) + \gamma(l\Omega -eV/\hbar) \\
\epsilon_l & =  & \epsilon(l\Omega +eV/\hbar) + \epsilon(l\Omega -eV/\hbar). 
\end{eqnarray*}
The functions $\gamma$ and $\epsilon$ are defined as
\begin{eqnarray*}
\gamma(\omega)  &  = & \frac{\pi \gamma_0}{\Omega} \int n_R(\omega') n_L(\omega+\omega') f(\omega')(1-f(\omega+\omega')) \, d\omega' \\ 
\epsilon(\omega) & = & \frac{\gamma_0}{\Omega} \mathcal{P} \int \frac{n_R(\omega') n_L(\omega'') f(\omega')(1-f(\omega''))}{\omega - \omega' + \omega''} \, d\omega' \,d\omega'' \, ,   
\end{eqnarray*}
where $\gamma_0= (R_K/R_T)\Omega/(2\pi)^2$, $f$ is the Fermi distribution and $n_L(\omega)$ ($n_R(\omega)$) is the dimensionless density of states of the left (right) contact of the junction that tend towards one at large $|\omega|$. As detailed for example in \cite{Catelani:2011cf,Catelani:2012cs}, the two functions $\gamma(\omega)$ and $\epsilon(\omega)$ can be fully determined from the knowledge of the junction $IV$ characteristic. The expressions above suppose that $n_L(\omega)$ and $n_R(\omega)$ are even function of $\omega$. Expressions in the general case are given in the SM. 

The regime of validity of the master equation is the weak tunneling limit, where the transmission of the barrier is sufficiently small in order to keep the contacts at equilibrium. The Born-Markov approximation also requires that the bath memory time is short compared to the resonator lifetime. Here, we are interested in the so-called strong coupling regime $\lambda \sim 1$, but we require that the coupling rate $\max \left( |w_{n,l}(\lambda)|^2 \right) \gamma_0$ is small compared to $\Omega$. We thus stay in the weak coupling regime as defined in quantum optics, which allows us to treat the coupling to the bath as a perturbation and to use the secular approximation. In terms of experimental parameters, we therefore require a tunnel junction with a large resistance $R_T \gg R_K$ such that $\gamma_0 \ll \Omega$ and a resonator with a high characteristic impedance $Z_c \sim R_K$ to obtain $\lambda \sim 1$. 

\section{Lamb shift}
Before considering specifically the features of equation (\ref{eq.mastereq}) for each type of junction, we review some of the properties of the Lamb shift and its consequences on the resonator spectrum. The Lamb shift operator $\sum_l \epsilon_l A_l A_l^\dagger$ is diagonal in the Fock state basis and shifts the Bohr frequency of the $|n\rangle$ state by
\begin{eqnarray}
\delta \omega_n & = & \sum_{l=-\infty}^\infty \epsilon_l \langle n | A_l A_l^\dagger | n \rangle \nonumber \\
& = & \sum_{l=1}^{\infty} \epsilon_{-l} |w_{n,l}(\lambda)|^2  + \sum_{l=0}^{n} \epsilon_{l} |w_{n-l,l}(\lambda)|^2
\label{eq.lambshift}
\end{eqnarray}
Each term in each sum can be interpreted as the contribution of a virtual process involving the exchange of $l$ photons between the resonator and the junction. In the weak coupling limit, we can expand the $w_{n,l}$ coefficients in powers of $\lambda$ and obtain the following expression for the frequency shift of the $n\rightarrow n+1$ transition :
\begin{multline*}
\delta \omega_{n+1} - \delta \omega_{n} = \gamma_0(\epsilon_1 + \epsilon_{-1} - 2\epsilon_0)\lambda^2 +  \gamma_0 \left[ n \epsilon_{2}  +  (n+2)\epsilon_{-2} \right. \\ \left. - (4n+2) \epsilon_{1} - (4n+6)\epsilon_{-1} + (6n+6) \epsilon_{0} \right] \frac{\lambda^4}{2} + O(\lambda^4).
\end{multline*}
The dominant term corresponds to a constant shift of all the transition frequencies, which is equivalent to a change of the resonator frequency $\delta \omega =  \gamma_0(\epsilon_1 + \epsilon_{-1} - 2\epsilon_0)\lambda^2$. The next order term gives rise to a nonlinear spectrum with a Kerr coefficient $U =  \gamma_0(\epsilon_{2} + \epsilon_{-2} - 4 \epsilon_{1} - 4 \epsilon_{-1} + 6 \epsilon_{0})\lambda^4/2$. Both the resonance shift and the nonlinearity depends on $V$ through the voltage dependence of $\epsilon_l$.  

As shown below, we find that $\epsilon(\omega)$ is an odd function of $\omega$ for the NIN junction therefore $\epsilon_l=-\epsilon_{-l}$. The resonance shift is thus or order $\lambda^4$ \cite{Dmytruk:2016uf} and seeing that the single photon loss coefficient $|w_{0,1}|^2$ is of order $\lambda^2$, we recover that the NIN junction acts as a purely dissipative element. The nonlinearity $U$ also cancels at order $\lambda^4$ and is of order $\lambda^6$.

The SIN and SIS junctions do not verify $\epsilon_l=-\epsilon_{-l}$, therefore the resonance shift and the single photon loss are both of order $\lambda^2$. One can estimate the order of magnitude of the resonance shift by supposing that the superconducting gap $\Delta$ is large compared to $\Omega$. We obtain that $\delta \omega \approx \gamma_0 (\Omega/\Delta) \lambda^2$ at zero bias and that it varies quadratically with $V$. If the bias voltage is kept inside the range where single photon loss are negligible, the variation of the resonance frequency with voltage can be on the order of the resonator linewidth even for relatively small values of $\lambda$. A similar calculation gives $U \approx \gamma_0 (\Omega/\Delta)^3 \lambda^4$ at zero bias. 

In the strong coupling regime, the contribution of higher powers of $\lambda$ cannot be neglected and the sum in $(\ref{eq.lambshift})$ must be computed numerically. We find that convergence is achieved by truncating the first sum after 20 terms when $\lambda^2 \leq 5$ and $n \leq 5$. In the SM, we show the evolution of the $0\rightarrow 1$ frequency shift as a function of $\lambda$ and voltage for each type of junction. The shift can be as large as a few $\gamma_0$ for the three types of junction. More importantly, the dependence of $\delta \omega_n$ with $n$ becomes highly nonlinear and we will use this feature to induce Zeno dynamics.  

\begin{figure}
\includegraphics{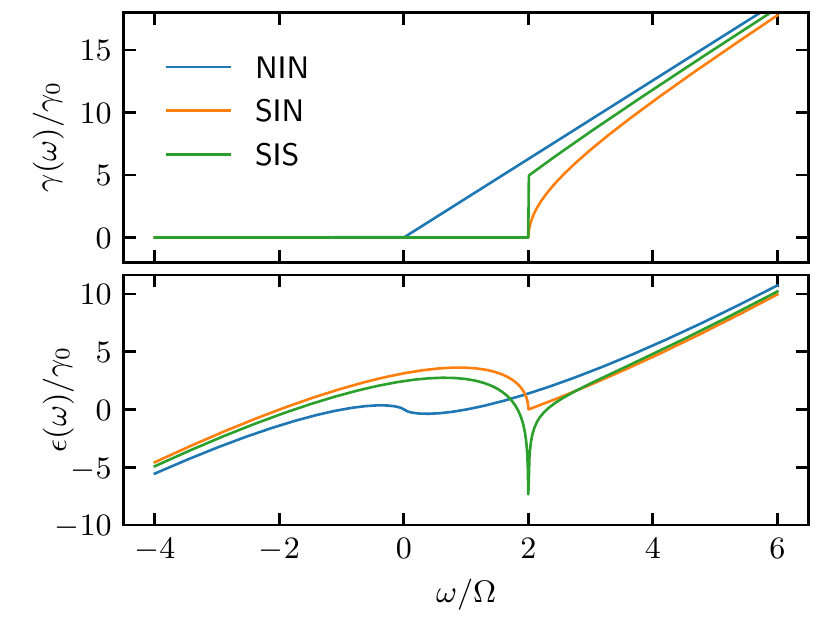}
\caption{Evolution of the real part $\gamma(\omega)$ and imaginary part $\epsilon(\omega)$ of the single sided Fourier transform of the memory function of the bath formed by the junction contacts at zero temperature. The function $\gamma_(\omega)$ is linked to the energy exchange (photon emission and absorption) between the resonator and the junction. The function $\epsilon_(\omega)$ describes virtual processes at the origin of the Lamb shift. For the SIN (SIS) junction, the gap is fixed to $2\hbar \Omega$ ($\hbar \Omega$). The abrupt rise of $\gamma(\omega)$ at the gap is accompanied by a dip in $\epsilon(\omega)$ as a consequence of the Kramers-Kronig relations.} 
\end{figure}

\section{NIN Junction : Single Photon Source}
In the case of a NIN junction ($n_R(\omega)=n_L(\omega)=1$), $\gamma(\omega)$ is proportional to the Bose-Einstein distribution $\gamma(\omega) = \pi (\gamma_0/\Omega)\, \omega/(1-\exp(-\beta \omega))$ with $\beta$ the inverse temperature. The integral defining $\epsilon(\omega)$ diverges but can be regularized with a high-energy cutoff. As shown in the SM, the diverging term corresponds to a global energy shift of the resonator spectrum and can therefore be removed. The final expression of $\epsilon(\omega)$ is independent of the cutoff and we obtain $\epsilon(\omega) = \gamma_0 (\omega/\Omega) \log |\omega/\Omega|$ at zero temperature (see figure 2). In the absence of external coherent drive, the energy shifts $\epsilon_l$ play no role and can be eliminated from the master equation, which simplifies to a rate equation between the diagonal terms of the density matrix. By choosing a bias voltage such that $2\hbar \Omega > |eV| > \hbar \Omega$, only jumps where the cavity gains at most one photon are allowed. If the coupling is set to $\lambda^2 = 2$, the jump from the Fock state $n=1$ to $n=2$ is forbidden and the resonator is populated with at most one photon. 

Figure 3 shows the evolution of $g^{(2)}(0)=\langle a^\dagger a^\dagger a a \rangle/\langle a^\dagger a \rangle^2$ as a function of $\lambda$ and $V$. As expected, we obtain $g^{(2)}(0) \approx 0$ when $\lambda^2 = 2$ and $2\hbar \Omega > |eV| > \hbar \Omega$. The residual value of $g^{(2)}(0)$ is given by the residual two photon pumping rate, which is on the order of $\pi \gamma_0 \exp(-\beta \Omega)$ and can be made very small in current experiments. The value of $g^{(2)}(0)=0$ is immune to voltage fluctuations and to extra single photon losses due to the coupling of the resonator to the outside world. The experimental difficulty relies in the design and the fabrication of a circuit where $\lambda$ can be tuned to $\sqrt{2}$ with sufficient precision. Because of the hierarchy $\gamma_l>\gamma_l'$ if $l>l'$, population inversion is not possible. The population of the $n=1$ Fock state tends towards $1/4$ when $V$ approaches $2\hbar \Omega$ from below at zero temperature. The Wigner function of the resonator mode thus remains positive at all $V$.

\begin{figure}
\includegraphics{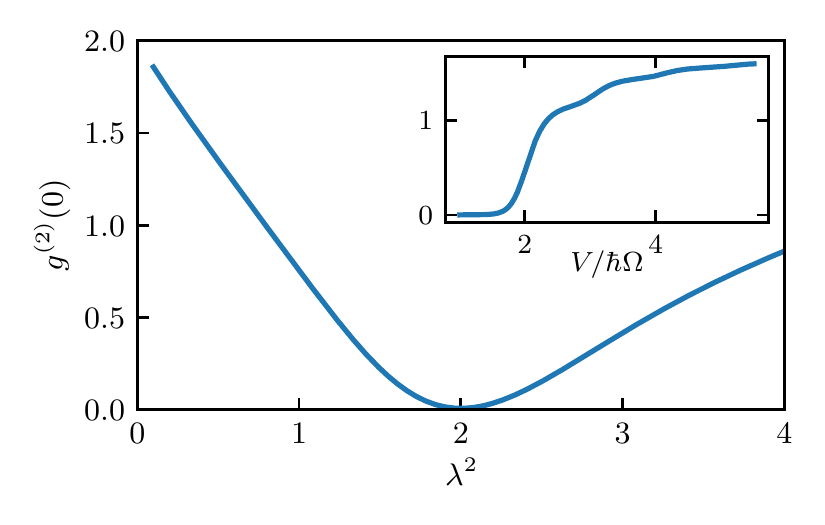}
\caption{Evolution of $g^{(2)}(0)$ for a resonator coupled to a NIN junction as a function of $\lambda$ at a bias voltage $eV=1.5\, \hbar \Omega$ and $\beta \hbar \Omega= 14$. A single photon source is obtained when $\lambda^2=2$ because the matrix element $\langle 1 | A_{1}^\dagger A_{1} | 1\rangle$ cancels, meaning that the resonator cannot absorb a photon when it is in the $|1\rangle$ state. The inset shows the evolution of $g^{(2)}(0)$ with voltage when $\lambda^2=2$. Increasing the voltage allows jumps where more than one photon are absorbed and $g^{(2)}(0)$ increases.}
\end{figure}

The effect of the Lamb shift can be probed by adding a coherent drive and looking, for example, at the resonator spectrum. As already mentioned, the first nonzero contribution to $\delta \omega$ in the weak coupling limit is of order $\lambda^4$ and we find $\delta \omega = \lambda^4 \gamma_0 (-4 \log 2 + 3(eV/\hbar\Omega)^2/2)$ at small bias and zero temperature. The prefactor can be rewritten $\lambda^4 \gamma_0 = (\pi/2) Z_c^2/(R_K R_T)\Omega$, which is proportional to the fine structure constant $\alpha=Z_0/(2R_K)$, where $Z_0$ is the impedance of free space, showing that this is a QED effect. The shift is small compared to the single photon loss rate $2\pi \gamma_0 \lambda^2$ and thus difficult to measure at small $\lambda$. But in the strong coupling regime, the shift of the $0\rightarrow 1$ transition can become on the order of $\gamma_0$ and comparable to the single photon loss. At a given $\lambda$, we observe that the shift increases from a negative value to a positive value as a function of bias and that it passes by an extremum in the region where it is negative (see SM). 
   
\section{SIN and SIS junction} 
We consider that the density of states of a superconducting contact is given by the BCS formula $\sqrt{\hbar^2 \omega^2/(\hbar^2 \omega^2-\Delta^2)}$, where $\Delta$ is the superconducting gap. In the case of the SIN junction, the two functions $\gamma(\omega)$ and $\epsilon(\omega)$ can be calculated analytically at zero temperature and their expressions are given in the SM. In the case of the SIS junction, we compute them numerically. As can be seen in figure 2, because of the superconducting gap, $\gamma(\omega)$ is zero for $\hbar \omega < \Deltat$, where $\Deltat=\Delta$ ($\Deltat=2\Delta$) for the SIN (SIS) junction. It rapidly increases when $\hbar \omega$ increases above $\Deltat$ because of the singularity in the BCS density of states. The effect is more pronounced in the SIS junction, where both contacts are superconducting. As a consequence, the jump rate $\gamma_l$ is zero as long as $|eV| \leq \Deltat - l\hbar\Omega$. The single photon source  described above for the NIN junction is easily adapted to the case of the SIS or SIN junction by choosing a bias voltage $\Deltat + \hbar\Omega < |eV| < \Deltat + 2 \hbar\Omega$ such that $\gamma_{-1}$ is non zero while $\gamma_{-2}=0$.

\begin{figure}
\includegraphics{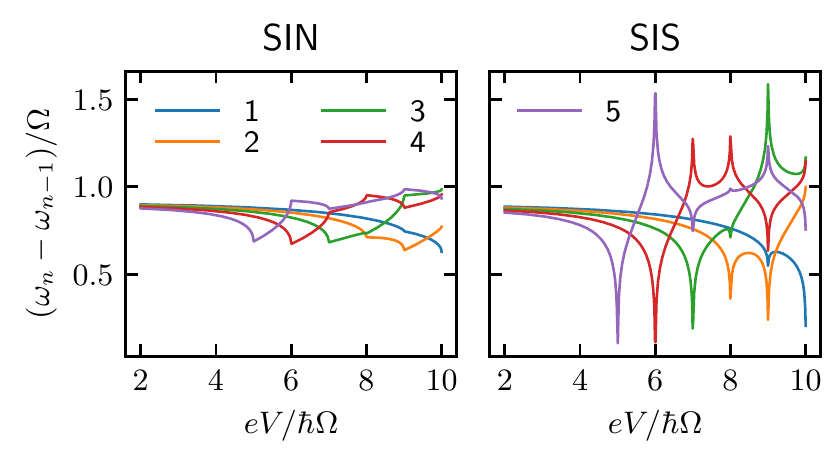}
\caption{Evolution of the transition frequencies $n-1 \rightarrow n$ as a function of voltage in the strong coupling regime ($\lambda^2=5$). Each resonance is indexed by the value of $n$ (upper state). The gap is chosen such that $\Deltat=10\hbar \Omega$ and $\gamma_0=\Omega/10$. The $n-1 \rightarrow n$ transition is strongly shifted when the bias voltage verifies $|eV| = \Deltat - l \hbar \Omega$ with $l \leq n$.}
\label{fig.spectrum}
\end{figure}

In relation to the abrupt change of the real part of the bath memory function, the imaginary part $\epsilon(\omega)$ has a dip at $\omega=\Deltat/\hbar$, which again is more marked in the SIS junction. The same dip appears in $\epsilon_l$ when $|eV| = \Deltat - l \hbar \Omega$. As a consequence, the $n-1 \rightarrow n$ transition is most shifted when $|eV| = \Deltat - l \hbar \Omega$ with $l \leq n$ (see figure \ref{fig.spectrum}). In the case where $|eV| = \Deltat - n \hbar \Omega$, the transition $n-1 \rightarrow n$ is shifted, whereas $\gamma_{l\leq n}=0$, meaning that the junction does not induce extra losses in the states $|1\rangle$ to $|n\rangle$. In the weak coupling regime, the resonance shift is proportional to $\gamma_0 \lambda^2  =(Z_c/R_T)\Omega/(4\pi)$, which can be interpreted as a classical effect due to the frequency dependence of the junction impedance. But the nonlinearity is proportional to $\gamma_0 \lambda^4$ and is of quantum origin, as discussed above for the frequency shift in the NIN junction case.

\subsection{Quantum Zeno Dynamics: Four photon cat state}
One can take advantage of the strong dependence with $V$ of $\gamma_l$ and $\epsilon_l$ in order to confine the dynamics of the resonator mode to the subspace spanned by the Fock states $|n\rangle$ with $n < n_c$ in the presence of a coherent drive. Two approaches can be followed to forbid the $n_c-1 \rightarrow n_c$ transition: one can introduce large losses in the $|n_c\rangle$ state, which is the original idea of the quantum Zeno dynamics (QZD) \cite{Facchi:2009up}, or equivalently one can apply a large energy shift to the $|n_c\rangle$ state, as realized for example in \cite{Signoles:2014tm}. The combination of both effects can also be used. The confinement is efficient if the loss rate or the energy shift is large compared to the driving $\eta$, which itself must be large compared to the residual nonlinearity and loss rate inside the confined subspace. Naively, one expects that the effect of losses is optimized for a voltage close to $|eV| = \Deltat - (n_c-1) \hbar \Omega$ and for a coupling $\lambda^2=n_c$ which maximizes $|w_{0,n_c}|^2$. This bias voltage also leads to a strong shift of the $n_c-1 \rightarrow n_c$ transition. Alternatively, $|eV| = \Deltat - n_c \hbar \Omega$ is also a good candidate in order to take advantage of the strong shift of $n_c-1 \rightarrow n_c$ without adding loss to $|n_c\rangle$. As shown below, this second option gives best results. 

In order to confirm that QZD can be obtained, we consider the restriction of the dynamics to the lowest $n_c=5$ Fock states, which contain at most 4 photons. We first simulate ideal QZD in a resonator and stop the evolution at the time where the population in $|4\rangle$ is maximal. The resulting state is a cat-like state, which we set as our target state. The Wigner distribution of the state is shown in figure 5. The goal is now to find the optimum $V$, $\lambda$, drive strength and detuning that maximizes the fidelity to the target state for the SIS and the SIN junction. For both junctions, the gap is chosen such that $\Deltat=10\hbar \Omega$. The exact value of the gap has little influence on the final result but it should be large enough so that the optimal bias values mentioned above exist. 

\begin{figure}
\includegraphics{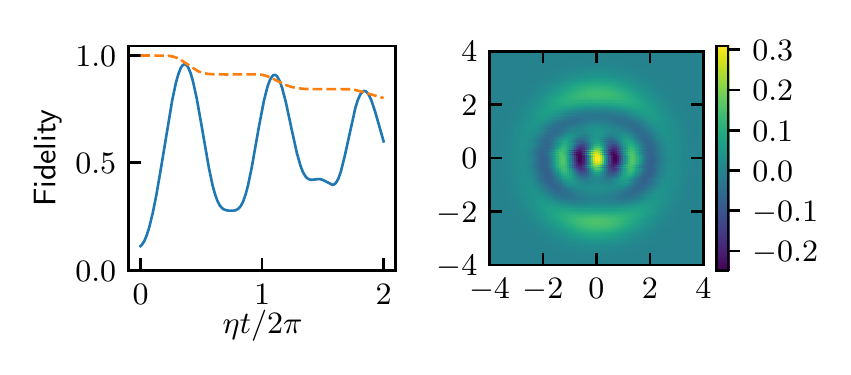}
\caption{Quantum Zeno dynamics in a driven resonator coupled to a SIS junction. The right graph shows the Wigner function of the target cat state. The left graph shows the time evolution of the fidelity to the target state (solid line) and the state purity (dashed line) starting from vacuum. The coupling is $\lambda^2=4.8$ and the bias $|eV|=\Deltat - 5\hbar\Omega$. The detuning and amplitude of the coherent drive are optimized to maximize the fidelity. The maximum fidelity is above 0.95 and the state purity decreases every time the state comes close to the boundary of the Zeno subspace.}
\end{figure}

We use the following procedure to optimize the parameters. We first look for the values of $V$ and $\lambda$ that maximize the ratio of the loss in $|5\rangle$ or the shift of $4 \rightarrow 5$ to the standard deviation of the transition frequencies $0 \rightarrow 1$, ..., $3 \rightarrow 4$. We simulate for these values the time evolution of $\rho$ starting from $\rho=|0\rangle\langle0|$. The quantum master equation used in the simulation corresponds to (\ref{eq.mastereq}) with an extra term $-i[ H_D ,\rho]$ to account for the presence of the drive, where $H_D=\eta(a+a^\dagger)-\delta a^\dagger a$. As a function of time, the fidelity passes through a maximum that we optimize by varying the detuning $\delta$ and amplitude $\eta$. We obtain a maximum fidelity of 0.92 (0.96) for the SIN (SIS) junction at an optimal point $|eV|=5\hbar \Omega$ and $\lambda^2=4.8$, which is the same for both junction. These high fidelities show that the dynamics is indeed very close to an ideal quantum Zeno dynamics. Figure 5 shows the time evolution of the fidelity in the case of the SIS junction with optimal parameters. For both junctions, the dependence of the fidelity with $\lambda$ is slow and lower coupling values can be used. For example, we obtain a fidelity of 0.95 for the SIS junction with $\lambda^2=2$. We also note that the optimal voltage bias is well below the gap, thus limiting the dc current to almost zero and avoiding any heating problem. The current through the junction is only due to rare quantum jumps, when the resonator leaves the confined subspace and $n_c$ (or more) photons are absorbed from the resonator allowing an electron to tunnel.    

\subsection{Qubit without Josephson Effect}

\begin{figure}
\includegraphics{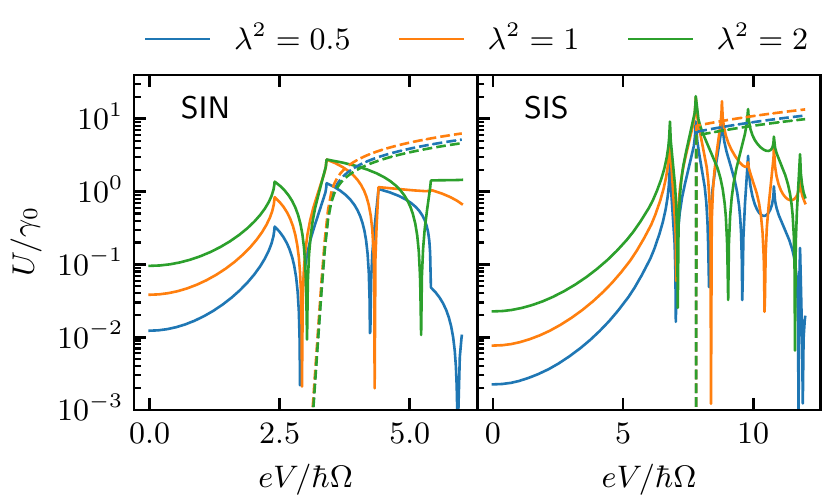}
\caption{Qubit nonlinearity with a strongly coupled SIN or SIS junction. Solid lines show the evolution of the qubit nonlinearity as a function of the voltage bias for different coupling values $\lambda$. Dashed lines show the single photon loss rate for the $|1\rangle$ qubit state.}
\end{figure}

Finally, we consider the realization of a qubit ($n_c=2$) using the Lamb shift induced nonlinearity. Figure 6 shows the evolution of the qubit nonlinearity $U$ and single photon loss as a function of the dc bias for different coupling parameters $\lambda$. The gap is chosen to be $\Delta = 4.4\,\hbar \Omega$, which corresponds to Aluminum and $\Omega=2\pi \times 10$\,GHz. For biases $|eV| < \Deltat- \hbar \Omega$, losses are negligible and the nonlinearity increases with $|V|$ reaching a maximum when $|eV| = \Deltat-2\hbar \Omega$. But increasing the bias voltage also increases the dependence of the qubit transition frequency on $V$, which is detrimental because it increases the phase noise induced by unavoidable bias fluctuations. Numerical simulations show that zero bias voltage maximizes the product of the qubit coherence time multiplied by the nonlinearity. 

We now suppose $V=0$ and consider the case of the SIN junction. We redefine the nonlinearity $U$ as $U=\delta\omega_2-2\delta\omega_1+\delta\omega_0$. In order to obtain $U=\Omega/20$, which is typical for a transmon qubit, we choose $\lambda^2=2$ and $\gamma_0=\Omega/2$. Losses introduced by the junction are completely negligible even at finite temperature and do not limit the qubit lifetime. We now estimate the qubit dephasing rate due to the bias voltage noise $\delta V$ for these parameters. The dependence of the qubit frequency is quadratic in $\delta V$ and we numerically find $\delta \omega_{01}/\gamma_0 \approx 5 \times 10^{-2} (e \delta V/\hbar \Omega)^2$. Because the junction is shunted by the inductance of the resonator, we expect that the low frequency voltage noise, which is responsible for the phase noise, will be dominated by flux noise and that charge noise should be negligible. Intrinsic flux noise with $1/f$ power spectral density leads to a rms voltage noise $\delta V \approx \delta \Phi f_c$ where $\delta \Phi \approx 1\,\mu\Phi_0$ and $f_c=1$\,GHz is a cutoff frequency \cite{Anton:2012en}. Using this value, we obtain a negligible dephasing rate on the order of $10^{-4}$\,s$^{-1}$. We foresee that the finite sub-gap conductance due to Andreev processes \cite{Rajauria:2008tp} or out-of-equilibrium processes \cite{Pekola:2010wr} will thus limit both the lifetime and the phase coherence time of the qubit, but possibly to high values.

In conclusion, we have shown that a resonator strongly coupled to a tunnel junction experiences both a state dependent Lamb shift and quantum jumps. This dependence can be used to realize a single photon source or to induce nonlinear coherent dynamics. Even though the apparition of a Lamb shift is a well known consequence of bath coupling, its engineering has not been given much consideration so far. Here, we clearly establish that this new facet of quantum bath engineering can be turned into an interesting resource. 

\begin{acknowledgments}
The authors would like to thank Christophe Mora and Fabien Portier for fruitful discussions. 
\end{acknowledgments}

\bibliography{biblio}

\end{document}


\title{Quantum dynamics of a microwave resonator strongly coupled to a tunnel junction: Supplementary Materials}

\author{J\'er\^ome Est\`eve}
\author{Marco Aprili}
\author{Julien Gabelli}
\affiliation{Laboratoire de Physique des Solides, CNRS, Universit\'e Paris-Sud, Universit\'e Paris-Saclay, Orsay, France}

\date{\today}

\maketitle

\section{Master Equation Derivation}
In the Born-Markov approximation, the master equation describing the evolution of the density matrix of a resonator coupled to a bath is 
\begin{equation}
\frac{d\rho}{dt} = -\frac{1}{\hbar^2} \int_0^\infty \Tr_e \left[ H_I(t) H_I(t-\tau) \rho(t) \otimes \rho_e-H_I(t)\rho(t)\otimes \rho_e H_I(t-\tau) + \hc \right] d\tau
\end{equation}
where $\rho_e$ is the density matrix of the bath, here the electrodes of the junction, and $H_I(t)$ is the coupling Hamiltonian between the resonator and the bath in the interaction picture. Here, $H_I(t)$ is the tunnel Hamiltonian
\begin{equation}
H_I(t) = \tilde{t} \sum_{k,q} c^\dagger_{R,k}(t) c_{L,q}(t) e^{i\lambda[a(t)+a^\dagger(t)]} e^{ieVt/\hbar} +\hc
\end{equation}
where $\tilde{t}$ is the barrier transmission and $c_{L,q}$ and $c_{R,k}$ are fermion operators associated to the left and right contacts of the junction. In order to apply the secular approximation, the coupling Hamiltonian must be decomposed as a sum of operators that couple levels with well defined energies. The displacement operator is thus decomposed as
\begin{eqnarray}
e^{i\lambda[a(t)+a^\dagger(t)]} & = & \sum_{n,l} | n+l \rangle \langle n+l | e^{i\lambda[a(t)+a^\dagger(t)]} | n \rangle \langle n | \\
& = & \sum_l A_l e^{i l \Omega t}
\end{eqnarray}
Using $A_l^\dagger = (-1)^l A_{-l}$, the coupling Hamiltonian can be rewritten
\begin{eqnarray}
H_I(t) & = & \tilde{t} \sum_l A_l e^{i l \Omega t} \left( \sum_{k,q} c^\dagger_{R,k}(t) c_{L,q}(t) e^{ieVt/\hbar} + (-1)^l c_{R,k}(t) c^\dagger_{L,q}(t) e^{-ieVt/\hbar} \right)\\
& = & \tilde{t} \sum_l A_l e^{i l \Omega t} B_l(t)
\end{eqnarray}
where $B_l(t)$ is a bath operator. The master equation can be rewritten in terms of $A_l$ and $B_l$ operators as
\begin{multline}
\frac{d\rho}{dt} = -\frac{\tilde{t}^2}{\hbar^2} \int_0^\infty \sum_{l,l'}  A_l A_{l'} \rho(t) \, e^{i l \Omega t} e^{i l' \Omega (t-\tau)} \, \Tr [B_l(t) B_{l'}(t-\tau) \rho_e] \\ 
- A_l \rho(t) A_{l'}  \, e^{i l \Omega t}e^{i l' \Omega (t-\tau)} \, \Tr [B_l(t) \rho_e B_{l'}(t-\tau) ] + \hc  \ d\tau
\end{multline}
The bath is supposed to be at stationary equilibrium and the expectation values of bath operators do not depend on $t$. The secular approximation, which consist in neglecting rapidly varying terms, thus leads to 
\begin{equation}
\frac{d\rho}{dt} = -\frac{\tilde{t}^2}{\hbar^2} \int_0^\infty \sum_{l} \left[ A_l A^\dagger_{l} \rho(t) K(\tau) e^{i l \Omega \tau} - A_l \rho(t) A^\dagger_{l} K(\tau) e^{i l \Omega \tau} + \hc \right] \ d\tau
\end{equation}
where the bath memory function $K(\tau)$ is defined as
\begin{equation}
K(\tau) =  \sum_{k,q} \langle c^\dagger_{R,k}c_{R,k}c_{L,q}c_{L,q}^\dagger\rangle e^{i(\omega_k-\omega_q)\tau}e^{i e V \tau /\hbar}  + \hc
\end{equation}
The average is taken over $\rho_e$ and the time dependence of the junction operators has been made explicit. We then define $\gamma_l$ and $\epsilon_l$ as the real and imaginary part of the single sided Fourier transform of $K(\tau)$ at frequency $l\Omega$
\begin{equation}
\frac{\tilde{t}^2}{\hbar^2} \int_0^\infty K(\tau) e^{i l \Omega \tau} = \gamma_l +i \epsilon_l
\end{equation}
and obtain the master equation given in the main text.

\subsection{Expressions of $\gamma_l$ and $\epsilon_l$}
Using the identity $\int_0^\infty \exp(i \omega t) \, dt= \pi \delta(\omega) + i \mathcal{P} \frac{1}{\omega}$, we obtain
\begin{eqnarray}
\gamma_l   & = & \pi \tilde{t}^2 n_0^2  \int n_R(\omega_k)n_L(\omega_q) f(\omega_k) (1-f(\omega_q)) \, \delta(l\Omega + \omega_k-\omega_q + eV/\hbar) \\ & & \ \  + n_R(\omega_k)n_L(\omega_q) (1-f(\omega_k)) f(\omega_q) \,  \delta(l\Omega - \omega_k  + \omega_q - eV/\hbar) \ d\omega_k d\omega_q  \\
\epsilon_l & = & \tilde{t}^2 n_0^2  \, \mathcal{P}  \int \frac{n_R(\omega_k)n_L(\omega_q) f(\omega_k) (1-f(\omega_q))}{l\Omega + \omega_k-\omega_q + eV/\hbar} + \frac{n_R(\omega_k)n_L(\omega_q) (1-f(\omega_k)) f(\omega_q)}{l\Omega - \omega_k  + \omega_q - eV/\hbar} \ d\omega_k d\omega_q \, ,
\end{eqnarray}
where $n_0$ is the electron density of states at the Fermi energy. Assuming that $n_L$ and $n_R$ are even functions of $\omega$ and introducing the tunnel conductance $1/R_T = (2\pi)^2 \tilde{t}^2 n_0^2/R_K$, we arrive at the expressions given in the main text.

\section{Lamb Shift at zero temperature}
We consider $n_L$ and $n_R$ to be even functions of $\omega$. At zero temperature, the expression of $\epsilon(\omega)$ is 
\begin{equation}
\epsilon(\omega) = \frac{\gamma_0}{\Omega} \mathcal{P} \int_0^\infty \frac{n_R(\omega') n_L(\omega'')}{\omega - \omega' - \omega''} \, d\omega' \,d\omega'' \, ,
\end{equation}
which diverges. We introduce a cutoff $A$ and limit the integration domain to $\omega'+\omega''<A$. Integrating over the sum and the difference of $\omega'$ and $\omega''$, the integral can be rewritten
\begin{equation}
\epsilon(\omega) = \frac{\gamma_0}{\Omega} \mathcal{P} \int_0^A \frac{1}{\omega - u} \left(  \int_0^u n_R(v) n_L(u-v) \, dv \right) \, du.
\end{equation}  

\subsection{NIN Junction}
In the case of the NIN junction, $n_L=n_R=1$ and we obtain
\begin{eqnarray}
\frac{\Omega}{\gamma_0} \epsilon(\omega) =  \mathcal{P} \int_0^A \frac{u}{\omega - u}  \, du & =  & -A -\omega \log(A-\omega) + \omega \log|\omega| \\
& \approx & -A -\omega \log A + \omega \log|\omega|,
\end{eqnarray}  
where the second line comes from the fact that the cut-off is large. Inserting this expression in the Lamb shift operator $\sum_l \epsilon_l A_l A^\dagger_l$, the two diverging terms that explicitly depend on the cutoff give two terms: a first one proportional to $A\sum_l A_l A^\dagger_l$ and a second one proportional to $\log A \sum_l l A_l A^\dagger_l$. Because $\sum_l A_l A^\dagger_l=\mathbb{I}$ and $\sum_l l A_l A^\dagger_l=-\lambda^2 \mathbb{I}$, these two terms correspond to a global energy shift that can be discarded. We finally obtain
\begin{equation}
\epsilon(\omega) =  \gamma_0 \frac{\omega}{\Omega}  \log |\frac{\omega}{\Omega}|.
\end{equation}   

\subsection{SIN and SIS Junction}
For the SIN junction, the integration over $v$ leads to 
\begin{eqnarray}
\frac{\Omega}{\gamma_0} \epsilon(\omega) =  \mathcal{P} \int_\Delta^A \frac{\sqrt{u^2-\Delta^2}}{\omega - u}  \, du . 
\end{eqnarray}
To regularize the integral, we subtract the integral corresponding to the NIN junction and compute 
\begin{eqnarray}
\frac{\Omega}{\gamma_0} \epsilon(\omega) =  \mathcal{P} \int_\Delta^A \frac{\sqrt{u^2-\Delta^2}-u}{\omega - u}  \, du , 
\end{eqnarray}
which does not diverge and can be calculated analytically. Adding back the contribution of the NIN junction without its diverging terms, we arrive at
\begin{equation}
\frac{\Omega}{\gamma_0} \epsilon(\omega) =
\begin{cases}
\sqrt{\omega^2-\Delta^2}  \log \left| \frac{\omega}{\Delta} + \sqrt{\frac{\omega^2}{\Delta^2}-1} \right| & \text{if $|\omega|>\Delta$} \\
2\sqrt{\Delta^2-\omega^2} \arctan  \sqrt{\frac{\Delta+\omega}{\Delta-\omega}} & \text{if $|\omega|<\Delta$} \\
\end{cases}
\end{equation}
In the case of the SIS junction, we could not obtain an analytic expression of $\epsilon(\omega)$. We follow the same procedure as for the SIN junction in order o compute numerically its value. We first subtract the contribution of the NIN junction to obtain a converging integral, which we evaluate numerically, and then add back the non diverging expression obtained for the NIN junction.

\subsubsection{Small coupling and large gap expansion at zero bias}
From the expansion of the $w_{n,l}$ coefficients, we obtain that $\delta \omega=\lambda^2\gamma_0(\epsilon_1 + \epsilon_{-1} - 2\epsilon_0)$. If the gap is large compared to the resonator frequency, we can approximate the expression in parenthesis by the second derivative of $\epsilon(\omega)$ and obtain at zero bias
\begin{equation}
\delta \omega=2 \lambda^2\gamma_0 \Omega \frac{d^2 \epsilon}{d\omega^2} = 4 \lambda^2\gamma_0 \Omega \, \mathcal{P} \int_0^\infty\frac{n_R(\omega') n_L(\omega'')}{(\omega - \omega' - \omega'')^3} \, d\omega' \,d\omega''
\end{equation}
Supposing the gap is large, the integral can be approximated as $-\int_\Delta^\infty 1/(\omega' + \omega'')^3 \, d\omega' d\omega''=-1/(4\Delta)$ and we obtain $\delta \omega \approx -\lambda^2\gamma_0 \Omega/\Delta$. For the SIN junction, an exact expansion gives $\delta \omega = -\pi\lambda^2\gamma_0 \Omega/(2\Delta)$. The same can be done for the non-linearity
\begin{eqnarray}
U=2 \lambda^4\gamma_0 \Omega^3 \frac{d^4 \epsilon}{d\omega^4} & = & 48 \lambda^4\gamma_0 \Omega^3 \, \mathcal{P} \int_0^\infty\frac{n_R(\omega') n_L(\omega'')}{(\omega - \omega' - \omega'')^5} \, d\omega' \,d\omega'' \\
& \approx & -48 \lambda^4\gamma_0 \Omega^3 \int_\Delta^\infty \frac{1}{( \omega' + \omega'')^5} \, d\omega' \,d\omega'' \\
& = & -\frac{\lambda^4\gamma_0 \Omega^3}{2\Delta^3}
\end{eqnarray}
For the SIN junction, the exact expansion gives $U=-3\pi \lambda^4\gamma_0 \Omega^3/(2\Delta^3)$.

\newpage
\subsection{Resonance shift as a function of bias and coupling}   
At large couplings, the energy shift of a given Fock state must be evaluated by computing numerically the sum in equation (2) of the main text. The figure below shows the evolution of the shift $\delta \omega_{01}$ of the $0\rightarrow 1$ transition as a function of the coupling $\lambda$ and the bias voltage for each type of junction. 
\begin{figure}[h!]
\includegraphics{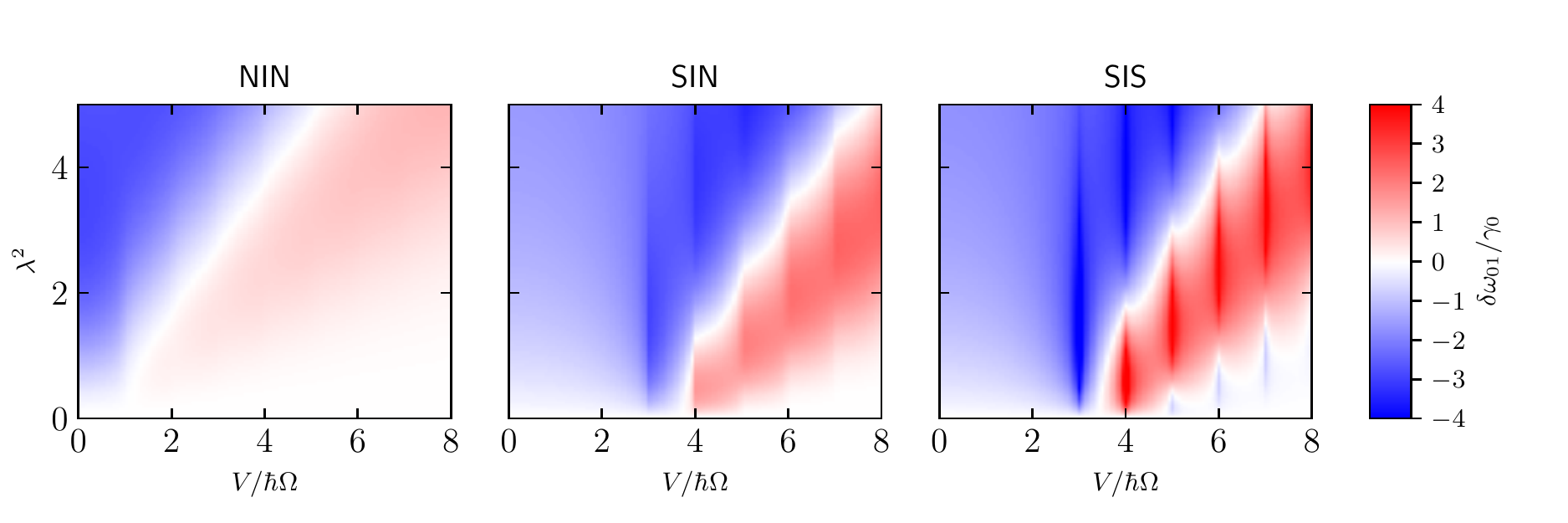}
\end{figure}